\newcommand{\beas}{\begin{eqnarray*}}
\newcommand{\eeas}{\end{eqnarray*}}

\newcommand{\diracslash}[1]{#1\llap{/\kern2pt}}

\newcommand{\be}{\begin{equation}}
\newcommand{\ee}{\end{equation}}
\newcommand{\bea}{\begin{eqnarray}}
\newcommand{\eea}{\end{eqnarray}}
\newcommand{\ba}[1]{\begin{array}{#1}}
\newcommand{\ea}{\end{array}}

\documentclass[preprint,prd,aps,floats,nofootinbib,floatfix]{revtex4}
\usepackage{graphicx}
\begin{document}
%

%

\title{Charmonium ground states in presence of strong magnetic fields}

\author{Amruta Mishra}
\email{amruta@physics.iitd.ac.in}
\affiliation{Department of Physics, Indian Institute of Technology, Delhi,
Hauz Khas, New Delhi -- 110 016, India}

\author{Pallabi Parui}
\email{pallabiparui123@gmail.com}
\affiliation{Department of Physics, Indian Institute of Technology, Delhi,
Hauz Khas, New Delhi -- 110 016, India}

\author{Ankit Kumar}
\email{ankitchahal17795@gmail.com}
\affiliation{Department of Physics, Indian Institute of Technology, Delhi,
Hauz Khas, New Delhi -- 110 016, India}

\author{Sourodeep De}
\email{sourodeepde2015@gmail.com}
\affiliation{Department of Physics, Indian Institute of Technology, Delhi,
Hauz Khas, New Delhi -- 110 016, India}

\begin{abstract}
The in-medium masses of the lowest S-wave charmonium states
(vector meson, $J/\psi$ and pseudoscalar meson, $\eta_c$) 
and P-wave charmonium states (scalar, $\chi_{c0}$ and axialvector
$\chi_{c1}$) are investigated in magnetized nuclear matter, within
the framework of QCD sum rule approach.
These are computed from the  medium modifications
of the scalar as well as twist--2 gluon condensates, obtained from
the medium modifications of a scalar dilaton field,
within a chiral effective model. The effects of the magnetic field,
isospin asymmetry and density on the masses of these charmonium
states have been investigated. 
The modifications of the masses of the P-wave charmonium
states ($\chi_{c0}$ and $\chi_{c1}$) are observed to be much 
larger as compared to those of the S-wave states, 
$J/\psi$ and $\eta_c$ within the QCD sum rule approach. 
The effects of the coupling of the spin
with the magnetic field are also investigated in the present work,
which result in the mixing of the spin zero charmonium state 
with the longitudinal component of the vector meson. This leads to
an increase (drop) in mass of the longitudinal $J/\psi$ ($\eta_c$)
for the S-wave states. The effects of the spin-magnetic field
interactions are observed to be dominant at high magnetic fields. 

\end{abstract}

\maketitle

\def\bfm#1{\mbox{\boldmath $#1$}}

\maketitle

\section{Introduction}
\label{intro}
The study of in-medium properties of hadrons, and,
more recently of the  heavy flavor hadrons 
\cite{Hosaka_Prog_Part_Nucl_Phys},
has been a topic of intense research, due to its relevance 
in ultra-relativistic heavy ion collision experiments.
The effects of magnetic field on the hadron properties
are also important to investigate, as strong magnetic fields
are known to be created in non-central high energy
heavy ion collision experiments \cite{HIC_mag}. The time
evolution of the magnetic field resulting from these 
high energy heavy ion collision experiments however 
is still an open question, which
needs the proper estimate of the electrical conductivity 
of the medium  as well as the solutions of the magnetohydrodynamic
equations.

In the QCD sum rule approach,
the mass modifications of the open heavy flavor (charm and bottom) mesons,
due to the presence of the light quark (antiquark) in these mesons,
arise from the medium modifications of the light quark condensates
\cite{open_heavy_flavour_qsr,Wang_heavy_mesons,arvind_heavy_mesons_QSR}.
On the other hand, the hidden heavy flavor mesons, 
e.g, the charmonium and bottomonium states
are modified due to the gluon condensates in the medium
within the QCD sum rule framework \cite{kimlee,klingl,amarvjpsi_qsr}.
In Ref. \cite{leeko}, the mass modifications of the charmonium
states have been studied using the leading order
QCD formula \cite{pes1} and
using the linear density approximation for the medium modification
of the gluon condensates in the nuclear medium. 
The QCD sum rule calculations for the charmonium states 
have been  generalized to finite temperatures \cite{moritalee},
where the medium modifications of these states have been
studied arising due to the temperature effects on the gluon 
condensates, extracted from lattice calculations.
The effects of magnetic fields have been investigated
on the properties of the open heavy flavor mesons,
e.g., D and B mesons  
\cite{Gubler_D_mag_QSR,machado_1,B_mag_QSR,dmeson_mag,bmeson_mag}
as well as the heavy quarkonium states
\cite{charmonium_mag_QSR,charmonium_mag_lee,charmonium_mag,upsilon_mag}.
The effects of the interaction of the spin with magnetic field
lead to mixing of the longitudinal component of the S=1 charmonium
state ($J/\psi$) with the S=0 state ($\eta_c$), resulting in 
a positive (negative) mass shift
in the longitudinal component of vector $J/\psi$ (pseudoscalar, $\eta_c$) 
\cite{charmonium_mag_QSR,charmonium_mag_lee,Suzuki_Lee_2017,Alford_Strickland_2013}. 

Using a chiral effective model, the mass modifications
of the open charm mesons have been studied,
arising from their interactions with the nucleons, hyperons 
and scalar mesons in the strange hadronic matter
\cite{amdmeson,amarindamprc,amarvdmesonTprc,amarvepja,DP_AM_Ds}.
The broken scale invariance of QCD is incorporated
in the chiral effective model through a scalar dilaton field
\cite{Schechter,paper3,hartree,kristof1}.
The in-medium masses of the charmonium states
have been investigated within the chiral effective model
\cite{amarvdmesonTprc,amarvepja},
by computing the scalar gluon condensate in the hadronic medium
from the medium modification of a scalar dilaton field.
This investigation showed a small drop
of the $J/\psi$ mass in the medium, whereas the masses of the excited
charmonium states are observed to have appreciable drop at high densities.
The chiral effective model has been generalized to the bottom sector
to study the in-medium masses of the open (strange) bottom mesons
\cite{DP_AM_bbar,DP_AM_Bs}
and the bottomonium states \cite{AM_DP_upsilon}.
The mass modifications of the 
open charm (bottom) as well as charmonium (bottomonium)
states have been used to compute the in-medium partial
decay widths of the charmonium (bottomonium) to 
$D\bar D$ ($B\bar B$) using a light quark pair model,
 namely $^3P_0$ model \cite{3p0,friman,amarvepja} 
as well as a field theoretic model for composite hadrons 
\cite{amspmwg,amspm_upsilon}. The effects of 
magnetic field have also been studied on the
masses of the open charm \cite{dmeson_mag},
open bottom \cite{bmeson_mag}, the charmonium
\cite{charmonium_mag} as well as the bottomonium
states \cite{upsilon_mag} in nuclear matter
including the effects of the anomalous magnetic moments 
of the nucleons. The decay widths of the charmonium to
$D\bar D$ in presence of strong magnetic fields
have also been studied using the $^3P_0$ model
\cite{charm_3p0_mag}
as well as using the field theoretic model for
composite hadrons \cite{charm_ft_mag}.

In the present investigation, we study the mass modifications of 
the S-wave states ($J/\psi$ and $\eta_{c}$) and P-wave states
($\chi_{c0}$ and $\chi_{c1}$) in the presence of strong magnetic fields, 
within a QCD sum rule framework \cite{klingl}, with the gluon condensates
in the hadronic medium calculated using the chiral effective model 
\cite{Schechter,paper3}. We calculate the contributions of 
the scalar gluon condensates, $\left\langle \frac{\alpha_{s}}{\pi}
G^a_{\mu\nu} {G^a}^{\mu\nu} \right\rangle$ and the twist-2 tensorial gluon 
operator, $\left\langle  \frac{\alpha_{s}}{\pi} G^a_{\mu\sigma}
{{G^a}_{\nu}}^{\sigma} \right\rangle $ upto dimension four \cite{klingl}
to investigate the masses of the S-wave states ($J/\psi$ and $\eta_c$)
as well as the P-wave states ($\chi_{c0}$ and $\chi_{c1}$) in nuclear matter
in the presence of a magnetic field. 
The effect of the interaction of spin to the external magnetic field
on the masses of the  S-wave states $J/\psi$ and $\eta_c$, 
has also been investigated, and this is observed to have 
dominant contribution at high magnetic fields.

The outline of the paper is as follows : In section II, we give a 
brief introduction to the chiral $SU(3)$ model used in the present
investigation, to calculate the in-medium gluon condensates needed 
to study the in-medium masses of charmonium ground states (vector $J/\psi$, 
pseudoscalar $\eta_c$, scalar $\chi_{c0}$ and axialvector $\chi_{c1}$)
using the QCD sum rule approach. The medium modifications of 
these charmonium states arise from the medium modification 
of the scalar and the twist-2 gluon condensates within the QCD sum rule
framework. In addition, the effects of the spin-magnetic field interaction
on the mass modifications for the S-wave charmonium
states ($J/\psi$ and $\eta_c$) are estimated. These lead to 
an increase (drop) in the mass of $J/\psi$ ($\eta_c$).
Section III discusses briefly the QCD sum rule approach used to 
calculate the masses of the S-wave and P-wave charmonium states.
In section IV, we discuss the results of the present investigation. 
Section V summarizes the findings of the present work. 
 
\section{ The hadronic chiral $SU(3) \times SU(3)$ model}
\label{sec:2}
In this section, we  briefly describe the chiral $SU(3)$ model
\cite{paper3} used to calculate the gluon condensates
in the magnetized nuclear matter. 
The in-medium masses of the charmonium states, 
$J/\psi$, $\eta_c$, $\chi_{c0}$ and $\chi_{c1}$, are then computed
from these gluon condensates, within the QCD sum rule framework.
The chiral SU(3) model is based on the nonlinear realization of chiral 
symmetry \cite{weinberg,coleman,bardeen} and broken scale invariance 
\cite{paper3,hartree,kristof1}. The broken scale invariance
is incorporated within the effective hadronic model 
with a scale breaking logarithmic potential
in terms of a scalar dilaton field, 
whose medium modifications yield the medium dependent gluon 
condensates.
The effective hadronic chiral Lagrangian density is given as
\be
{\cal L} = {\cal L}_{kin} + {\cal L}_{BM}
          + {\cal L}_{vec} + {\cal L}_0 +
{\cal L}_{scalebreak}+ {\cal L}_{SB}+{\cal L}_{mag}.
\label{genlag} \ee 
In the above Lagrangian density, the first term ${\cal L}_{kin}$ 
corresponds to the kinetic energy terms of the baryons and the mesons.
${\cal L}_{BM}$ is the baryon-meson interaction term,
${\cal L}_{vec}$  corresponds to the interactions of the vector 
mesons, ${\cal L}_{0}$ contains the meson-meson interaction terms, 
${\cal L}_{scalebreak}$ is a scale invariance breaking logarithmic 
potential given in terms of a scalar dilaton field \cite{heide1}, 
$ {\cal L}_{SB} $ is the explicit chiral symmetry
breaking term, and ${\cal L}_{mag}$ is the contribution 
from the magnetic field, given as 
\cite{broderick1,broderick2,Wei,mao,dmeson_mag,bmeson_mag,charmonium_mag,upsilon_mag}
\be 
{\cal L}_{mag}=-{\bar {\psi_i}}q_i \gamma_\mu A^\mu \psi_i
-\frac {1}{4} \kappa_i \mu_N {\bar {\psi_i}} \sigma ^{\mu \nu}F_{\mu \nu}
\psi_i
-\frac{1}{4} F^{\mu \nu} F_{\mu \nu},
\label{lmag}
\ee
where, $\psi_i$ is the field operator for the $i$-th baryon 
($i=p,n$, for nuclear matter, as considered in the present work),
and the parameter $\kappa_i$ in the second term in equation (\ref{lmag}) 
is related to the anomalous magnetic moment of the $i$-th baryon
\cite{broderick1,broderick2,Wei,mao,amm,VD_SS,aguirre_fermion}.
The values of $\kappa_p$ and  $\kappa_n$ are given as 
$3.5856$ and $-3.8263$ respectively, which are the values
of the gyromagnetic ratio corresponding to the 
anomalous magnetic moments of the proton and 
the neutron respectively.

Within the chiral SU(3) model used in the present investigation,
the scalar gluon condensate 
$ \langle \frac{\alpha_s}{\pi}  {G^a}_{\mu\nu} G^{a\mu\nu} \rangle $,
as well as the twist-2 gluon operator,
$ \langle \frac{\alpha_s}{\pi}  {G^a}_{\mu\sigma} 
{{G^a}_\nu}^{\sigma} \rangle $,
are simulated by the scalar dilaton field, $\chi$. These are
obtained from the energy momentum tensor 
\begin{eqnarray}
T_{\mu \nu}=(\partial _\mu \chi) 
\Bigg (\frac {\partial {{\cal L}_\chi}}
{\partial (\partial ^\nu \chi)}\Bigg )
- g_{\mu \nu} {\cal L}_\chi,
\label{energymom}
\end{eqnarray}
derived from the Lagrangian density for the dilaton field 
\cite{amarvjpsi_qsr}.
The energy momentum tensor in QCD, accounting for the current quark masses,
 can be written as
\cite{moritalee,cohen} 
\begin{eqnarray}
T_{\mu \nu}=-ST({G^a}_{\mu\sigma} {{G^a}_\nu}^{ \sigma})
+ \frac {g_{\mu \nu}}{4} 
\Bigg (\sum_i m_i \bar {q_i} q_i+ \langle \frac{\beta_{QCD}}{2g} 
G_{\sigma\kappa}^{a} {G^a}^{\sigma\kappa} \rangle \Bigg)
\label{energymomqcd}
\end{eqnarray}
where the first term is the symmetric traceless part and second term
is the trace part of the energy momentum tensor.
Writing 
\begin{eqnarray}
\langle \frac {\alpha_s}{\pi}{G^a}_{\mu\sigma} {{G^a}_\nu}^{\sigma} \rangle
=\Big (u_\mu u_\nu - \frac{g_{\mu \nu}}{4} \Big ) G_2,
\label{twist2g2}
\end{eqnarray}
where $u_\mu$ is the 4-velocity of the nuclear medium,
taken as $u_\mu =(1,0,0,0)$, we obtain the energy momentum tensor
in QCD as 
\begin{eqnarray}
T_{\mu \nu}=-\Big (\frac{\pi}{\alpha_s}\Big)\Big (u_\mu u_\nu - 
\frac{g_{\mu \nu}}{4} \Big ) G_2
+ \frac {g_{\mu \nu}}{4} 
\Bigg (\sum_i m_i \bar {q_i} q_i+ \langle \frac{\beta_{QCD}}{2g} 
G_{\sigma\kappa}^{a} {G^a}^{\sigma\kappa}\rangle \Bigg).
\label{energymomqcdg2}
\end{eqnarray}

Equating the energy-momentum tensors given by
equations (\ref{energymom}) and (\ref{energymomqcdg2}) and
multiplying by $(u^\mu u^\nu -\frac {g^{\mu \nu}}{4})$
and $g^{\mu \nu}$ lead to the
expressions for $G_2$ (related to the twist--2 gluon condensate) 
and the scalar gluon condensate as \cite{amarvjpsi_qsr}
\begin{eqnarray}
G_2 &=&  \frac{\alpha_s}{\pi}
\Bigg [-(1-d+4 k_4)(\chi^4-{\chi_0}^4)-\chi ^4 {\rm {ln}}
\Big (\frac{\chi^4}{{\chi_0}^4}\Big )\nonumber \\ 
& + & \frac {4}{3} d\chi^{4} {\rm {ln}} \Bigg (\bigg( \frac {\left( \sigma^{2} 
- \delta^{2}\right) \zeta }{\sigma_{0}^{2} \zeta_{0}} \bigg) 
\bigg (\frac {\chi}{\chi_0}\bigg)^3 \Bigg ) \Bigg ],
\label{g2approx}
\end{eqnarray}
and,
\begin{equation}
\left\langle  \frac{\alpha_{s}}{\pi} {G^a}_{\mu\nu} {G^a}^{\mu\nu} 
\right\rangle =  \frac{8}{9} \Bigg [(1 - d) \chi^{4}
+\left( \frac {\chi}{\chi_{0}}\right)^{2} 
\left( m_{\pi}^{2} f_{\pi} \sigma
+ \big( \sqrt {2} m_{k}^{2}f_{k} - \frac {1}{\sqrt {2}} 
m_{\pi}^{2} f_{\pi} \big) \zeta \right) \Bigg ]. 
\label{chiglu}
\end{equation}
The scalar gluon condensate also depends on the 
scalar fields, $\sigma$ and $\zeta$, as we have included
the term due to the finite quark masses, in
the energy momentum tensor in QCD.
The in-medium masses of the charmonium states are 
modified due to the scalar gluon condensate and the twist-2
gluon operators (through the expression $G_2$). 

\section{QCD sum rule approach and in-medium charmonium masses}
\label{sec:3}
In the present section, we shall use the medium modifications
of the gluon condensate, calculated from the dilaton field
in the chiral effective model, to compute the masses of the
lowest charmonium states in isospin asymmetric 
nuclear matter in the presence of a magnetic field.
The masses of the S-wave charmonium states, $J/\psi$ and $\eta_c$,
have been studied in Ref. \cite{amarvjpsi_qsr}
for the case of zero magnetic field. Using the QCD sum rule 
approach \cite{klingl}, the in-medium mass squared of the charmonium 
state of type $i$  ($i$=Vector, Pseudoscalar, Scalar, Axialvector), 
${m^*_i}^2$, 
is given by
\begin{equation}
{m^*_i}^{2} \simeq \frac{M_{n-1}^{i} (\xi)}{M_{n}^{i} (\xi)}
 - 4 m_{c}^{2} \xi
\label{masscharm}
\end{equation}
where $M_{n}^{i}$ is the $n$th moment of the meson $i$ and $\xi$ is the 
normalization scale.  Using operator product expansion, the moment 
$M_{n}^{i}$ can be written as \cite{klingl}
\begin{equation}
M_{n}^{i} (\xi) = A_{n}^{i} (\xi) \left[  1 + a_{n}^{i} (\xi) \alpha_{s} 
+ b_{n}^{i} (\xi) \phi_{b} + c_{n}^{i} (\xi) \phi_{c} \right],
\label{moment}  
\end{equation}
where $A_n^i(\xi)$, $a_n^i(\xi)$, $b_n^i (\xi)$ and $c_n^i (\xi)$
are the Wilson coefficients. The common factor $A_{n}^{i}$ results 
from the bare loop diagram. The coefficients $a_{n}^{i}$ take into 
account perturbative radiative corrections, while the coefficients 
$b_{n}^{i}$ are associated with the scalar gluon condensate term
\begin{equation}
\phi_{b} = \frac{4 \pi^{2}}{9} \frac{\left\langle \frac{\alpha_{s}}{\pi} 
G^a_{\mu \nu} {G^a}^{\mu \nu} \right\rangle }{(4 m_{c}^{2})^{2}}, 
\label{phib} 
\end{equation}
where, the scalar gluon condensate is given by equation (\ref{chiglu}).
The coefficients $A_{n}^{i}, a_{n}^{i},$ and $b_{n}^{i}$ are listed in 
Ref.\cite{rein}. The coefficients $c_n^i$ are associated with the 
value of $\phi_{c}$, which gives the contribution from twist-2 gluon 
operator and is given as
\begin{equation}
\phi_{c} =  \frac{4 \pi^{2}}{3(4 m_{c}^{2})^{2}}G_2,  
\label{phicglu}
\end{equation}
where $G_2$ is given by equation (\ref{g2approx}).
The Wilson coefficients, $c_{n}^{i}$ in the vector channel (for $J/\psi$) and 
the pseudoscalar channel (for $\eta_{c}$) can be found in Ref. \cite{klingl}.
Using a background field technique, the coefficients $c^i_n$'s have been
calculated 
for the P-wave charmonia, $\chi_{c0}$ and $\chi_{c1}$ 
in Ref. \cite{Song_Lee_Morita_Pwave_charmonia}. 
The parameters $m_{c}$ and $\alpha_{s}$ are the running charm quark mass 
and running coupling constant and are $\xi$ dependent \cite{rein}. 
In the next section, we present and discuss the results of our present 
work of the investigation of the in-medium masses of the S-wave states 
$J/\Psi$ and $\eta_c$, as well as, P-wave states $\chi_{c0}$ and
$\chi_{c1}$, in isospin asymmetric nuclear matter in the presence 
of magnetic field. The spin-magnetic field interaction 
leads to interaction between the longitunal component of the
spin 1 to the spin 0 state. The effects of the spin-magnetic field
interaction has been taken into account for the S-wave charmonium
states $J/\psi$ and the pseudoscalar meson $\eta_c$ 
\cite{charmonium_mag_QSR,charmonium_mag_lee,Suzuki_Lee_2017,Alford_Strickland_2013}. 
This interaction is observed to give mixing 
between the longitudinal component of the spin one (triplet) 
charmonium state, $J/\psi$ and the spin zero (singlet) pseudoscalar 
meson, $\eta_c$. 
In the present work, we estimate the modifications arising 
from the additional interaction of the spin with the magnetic field, 
for the S-wave charmonium masses, $J/\psi$ (spin 1) and $\eta_c$
(spin 0).
This leads to the effective masses for the longitunal $J/\psi$ and $\eta_c$ 
(accounting for the in-medium charmonium masses calculated 
within the QCD sum rule approach as given by equation (\ref{masscharm}),
along with the shift due to the spin-magnetic field interaction) 
to be given as \cite{Alford_Strickland_2013}
\begin{eqnarray}
m_{J/\psi}^{eff}=m^*_{J/\psi}+\Delta m_{sB},\;\; 
m_{\eta_c}^{eff}=m^*_{\eta_c}-\Delta m_{sB},
\label{effmass_swave}
\end{eqnarray}
In the above, the contribution due to the spin-magnetic field
interaction is given as,
\begin{equation}
\Delta m_{sB}=\frac{\Delta M}{2}\Bigg((1+\chi_{sB}^2)^{1/2}-1\Bigg),\;\; 
{\rm with }\;\;\chi_{sB}=\frac{2g \mu_c B}{\Delta M}, 
\label{msB}
\end{equation}
where $\mu_c=(\frac {2}{3}e)/(2m_c)$ as the (charm) quark Bohr magneton,
$\Delta M= m^*_{J/\psi} -m^*_{\eta_c}$, and $g$ is chosen to be 2,
thereby ignoring the anomalous magnetic moment effects
for the charm quark (antiquark).
In the expression of $\chi_{sB}$ given by equation (\ref{msB}),
$m_c$ is taken to be 1.8 GeV. 
The effects of the interaction of the spin with the magnetic field
is observed to lead to a rise (drop) in the longitudinal $J/\psi$
($\eta_c$) and these effects dominate over the mass modifications
arising within the QCD sum rule approach in presence of strong 
magnetic fields. 

\begin{figure}
\includegraphics[width=18cm,height=18cm]{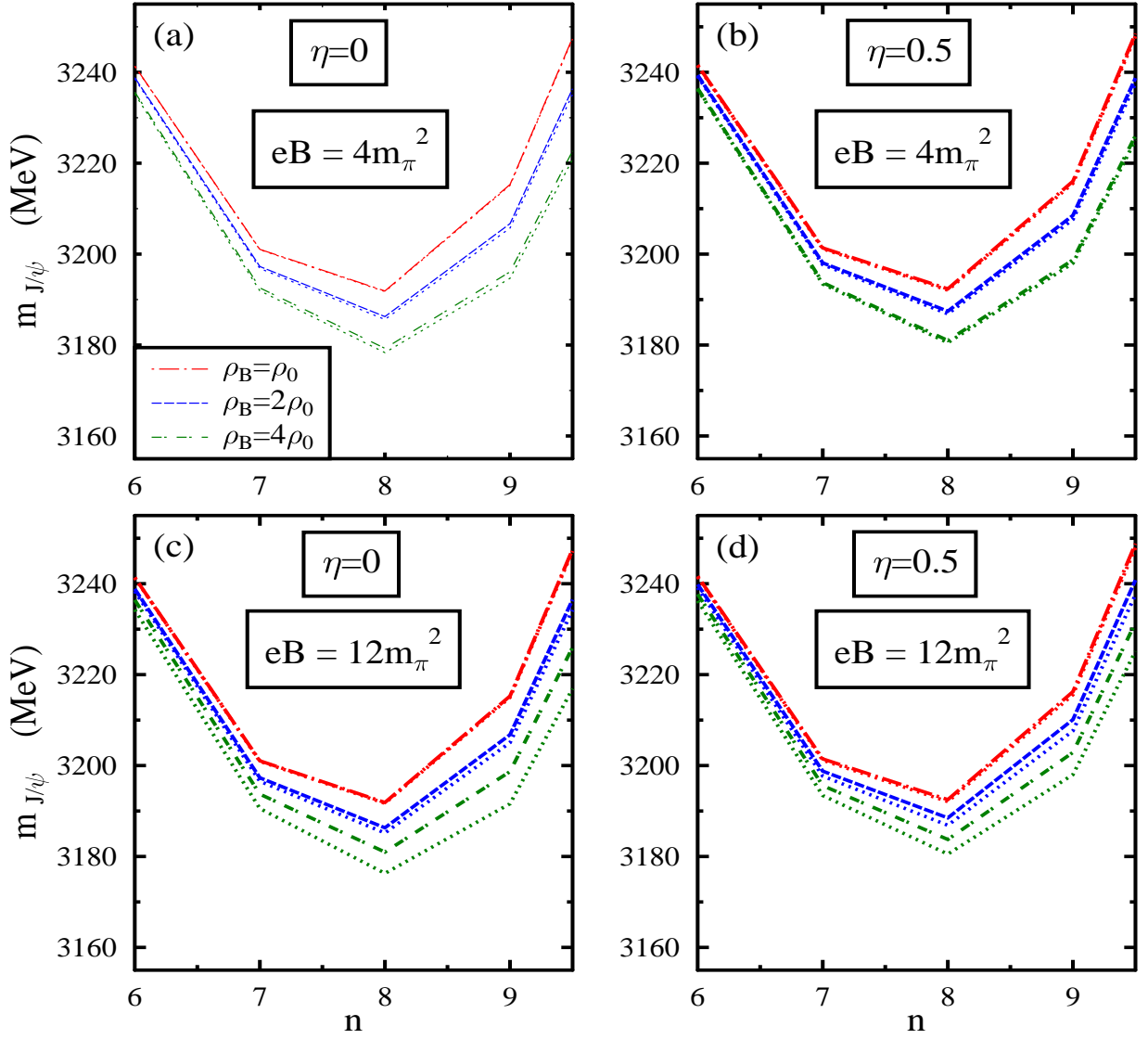}
\caption{(Color online) The mass of the vector charmonium state,
$J/\psi$ plotted as a function 
of n at given values of baryon densities, 
isospin asymmetry parameter and magnetic fields.
The value of $\xi$ is chosen to be 1, which yields 
the vacuum mass of $J/\psi$ to be 3196.2 MeV.} 
\label{mjpsi_mag_qsr}
\end{figure}

\begin{figure}
\includegraphics[width=18cm,height=18cm]{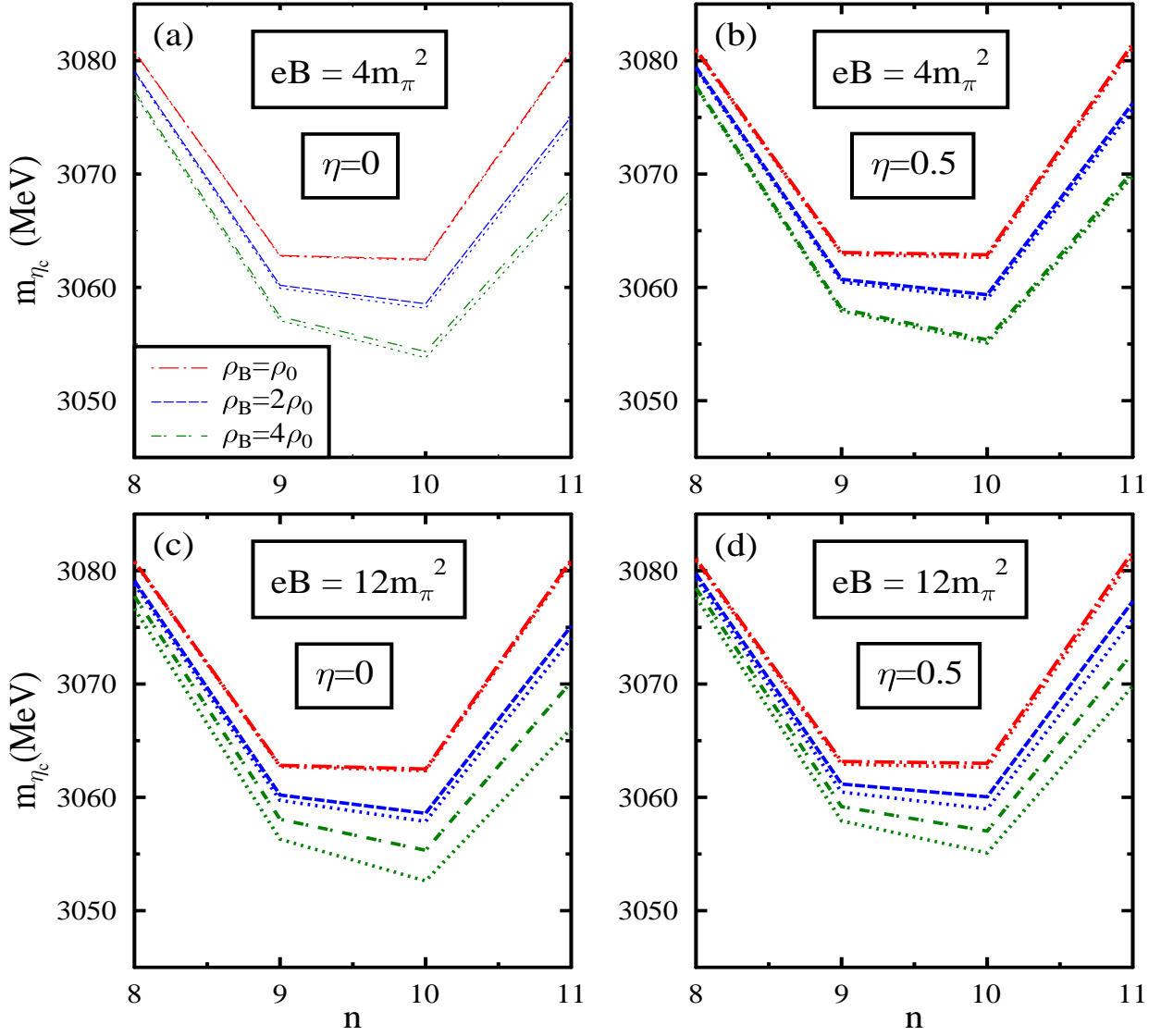}
\caption{(Color online) The mass of the  pseudoscalar 
charmonium state, $\eta_c$ plotted as a function 
of n at given values of baryon densities, 
isospin asymmetry parameter and magnetic fields.
The value of $\xi$ is chosen to be 1, which yields 
the vacuum mass of $\eta_c$ to be 3066.56 MeV.} 
\label{metac_mag_qsr}
\end{figure}

\begin{figure}
\includegraphics[width=18cm,height=18cm]{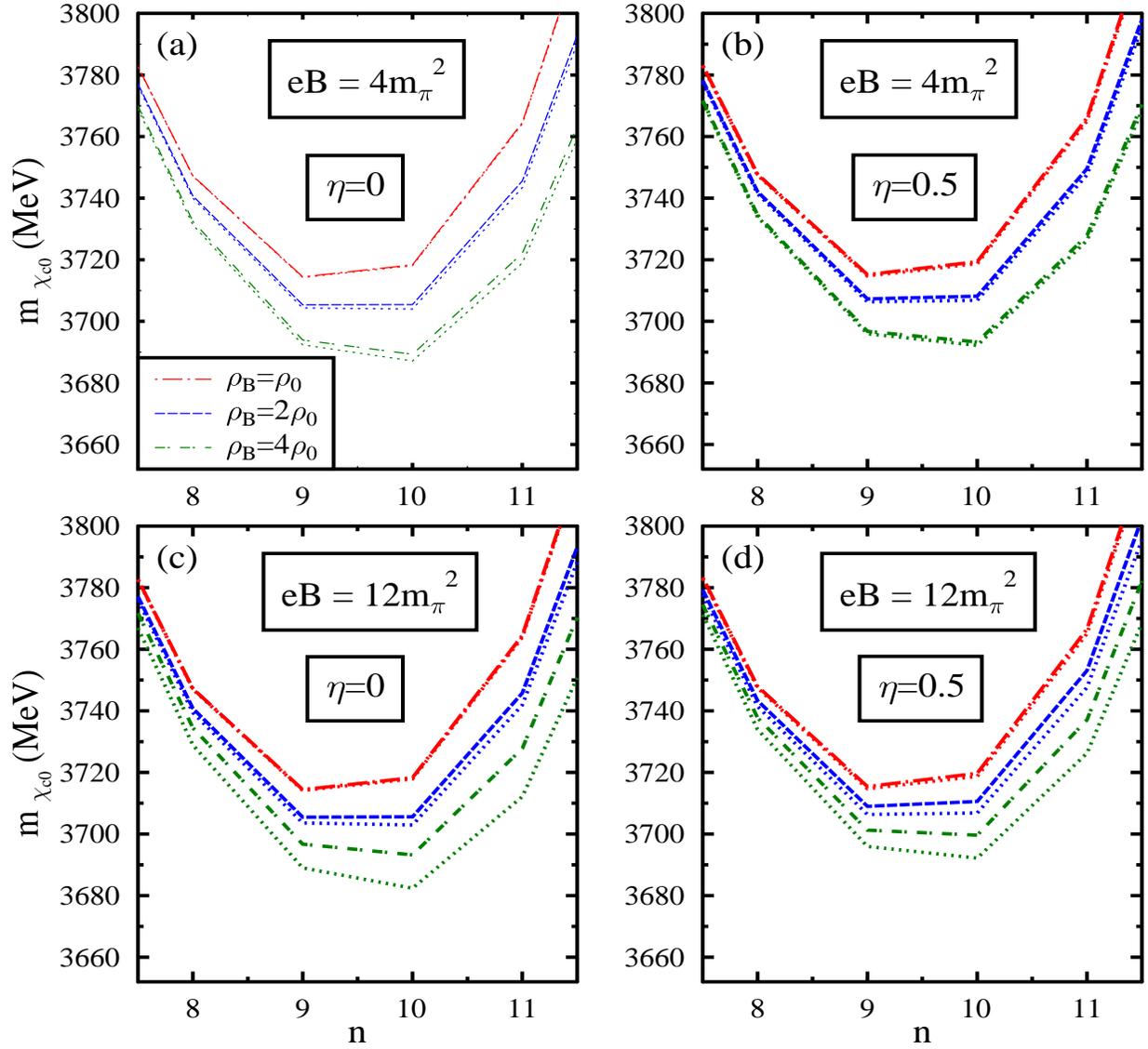}
\caption{(Color online) The mass of the scalar charmonium state, 
$\chi_{c0}$ plotted as a function 
of n for $\xi=2.5$, at given values of baryon densities, 
isospin asymmetry parameter and magnetic fields.
The value of $\xi$ is chosen to be 2.5, which yields 
the vacuum mass of $\chi_{c0}$ to be 3720.6264 MeV.} 
\label{mscalar_mag_qsr_xi2.5}
\end{figure}

\begin{figure}
\includegraphics[width=18cm,height=18cm]{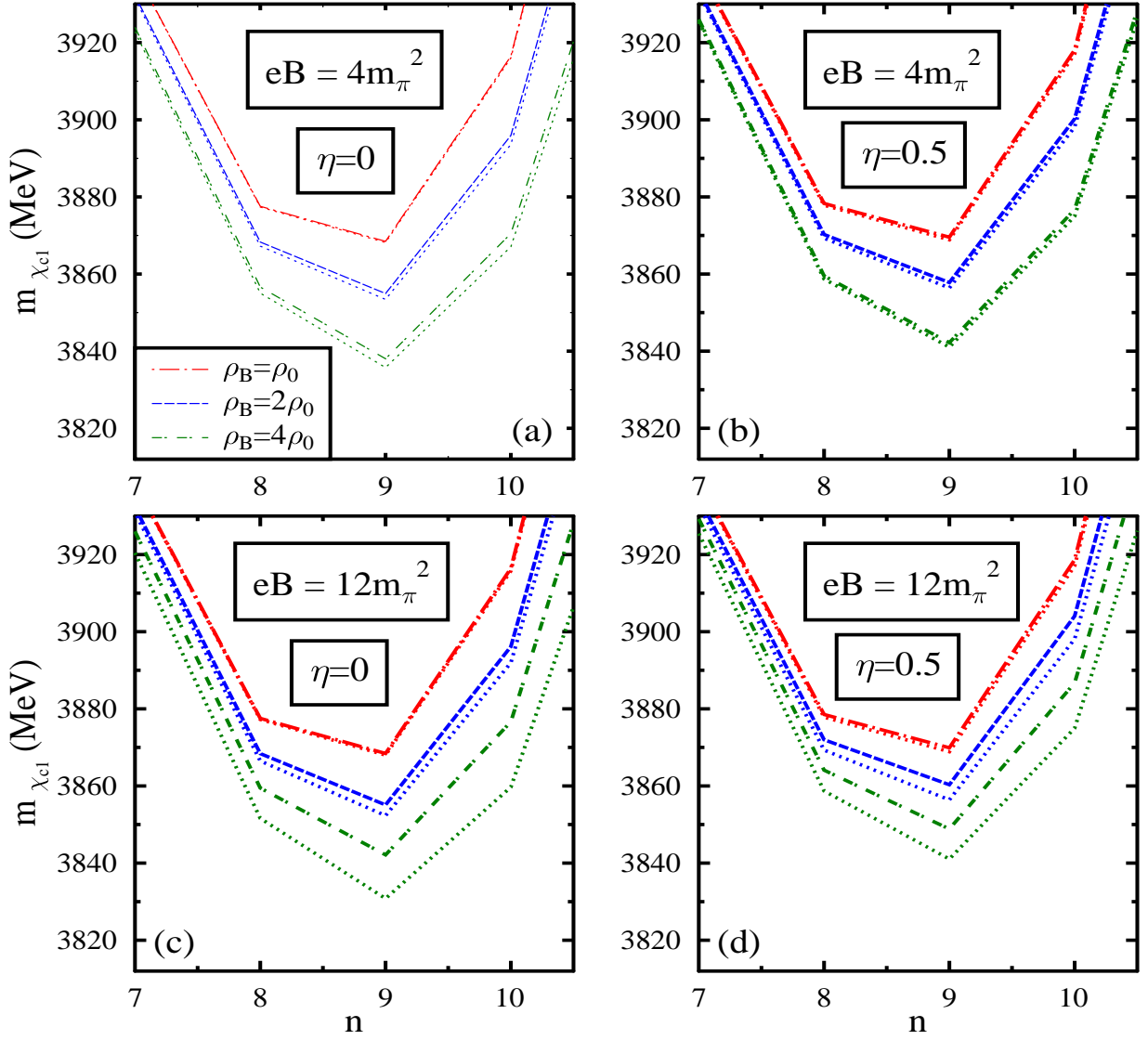}
\caption{(Color online) The mass of axial vector charmonium state 
$\chi_{c1}$ plotted as a function 
of n for $\xi=2.5$, at given values of baryon densities, 
isospin asymmetry parameter and magnetic fields.
The value of $\xi$ is chosen to be 2.5, which yields 
the vacuum mass of $\chi_{c1}$ to be 3877.77 MeV.} 
\label{maxialvec_mag_qsr_xi2.5}
\end{figure}

\section{Results and Discussions}
\label{sec:4}
In this section, we investigate the in-medium masses 
of the S-wave charmonium states $J/\psi$ ($^3S_1$) 
and $\eta_c$ ($^1S_0$) mesons 
and P-wave states $\chi_{c0}$ ($^3P_0$) and $\chi_{c1}$ ($^3P_1$)
in nuclear matter in the presence of strong magnetic fields
using the QCD sum rule approach \cite{klingl,kimlee,moritalee,amarvjpsi_qsr}.
The scalar and the twist-2 gluon condensates are calculated
within the chiral effective model, from the medium modification
of a scalar dilaton field, $\chi$, which mimics the gluon
condensates of QCD.
The in-medium masses of $J/\psi$ and $\eta_c$ have been calculated
for zero magnetic field within 
the QCD sum rule framework using the gluon condensates 
calculated within the chiral effective model \cite{amarvjpsi_qsr}. 
In the present work, the effects of magnetic field
are studied on the masses of the S- and P-wave charmonium states, 
and the effects of the interaction of the spin and magnetic field
on the masses of the $J/\psi$ and $\eta_c$ are investigated. 

\begin{table}
\begin{tabular}{|c|c|c|c|c|c|}
\hline
\multicolumn{2}{|c|}{
} & \multicolumn{2}{|c|}{$ \eta=0 $} 
&  \multicolumn{2}{|c|}{$ \eta=0.5 $}\\
\cline{3-6}
\multicolumn{2}{|c|}
 {${eB}/{m_\pi^2}$}
 &  $  \rho_B=\rho_0 $ & 
$ \rho_B=2\rho_0 $ & 
$ \rho_B=\rho_0 $
 & $ \rho_B=2\rho_0 $ 
\\
\hline
& (a) 
&  -4.424 & -10.574
&  -4.126 & -9.332
 \\
\cline{2-6}
{4}
& (b) 
 & -4.3  & -9.94
 & -3.838   & -8.75
      \\
\cline{2-6}
& (c) 
 & +1.748  &  -3.817  
& +2.2076  & -2.6486    
      \\
\hline
&  (a)
 &   -4.496   & -11.025
 &  -4.126 & -9.332
\\
\cline{2-6}
{ 12 }
& (b)
 & -4.316  & -9.8806
 & -3.689 & -7.7056
\\
\cline{2-6}
& (c) 
 & +38.506  &  +33.27  
 & +39.107  &  +35.3 
\\
\hline
\end{tabular}
\vskip 0.1in
\caption{Mass shifts of $J/\psi$ in magnetized nuclear
matter for densities of $\rho_0$ and 2$\rho_0$, asymmetric parameter,
$\eta$=0 and 0.5 and for magnetic fields, $eB/{m_\pi^2}$ as 4 and 12. 
These are tabulated without and with the contributions 
from anomalous magnetic moments of the nucleons 
in (a) and (b) respectively. The effects from the interaction
of spin to magnetic field are also taken into account in (c). }
\label{table1}
\end{table}

\begin{table}
\begin{tabular}{|c|c|c|c|c|c|}
\hline
\multicolumn{2}{|c|}{
} & \multicolumn{2}{|c|}{$ \eta=0 $} 
&  \multicolumn{2}{|c|}{$ \eta=0.5 $}\\
\cline{3-6}
\multicolumn{2}{|c|}
 {${eB}/{m_\pi^2}$}
 &  $  \rho_B=\rho_0 $ & $ \rho_B=2\rho_0 $ & 
$ \rho_B=\rho_0 $ & $ \rho_B=2\rho_0 $ \\
\hline
& (a) 
& -4.167 & -8.406
& -3.912 & -7.585
 \\
\cline{2-6}
{ 4}
& (b) 
& -4.07 & -8
& -3.68 & -7.21
 \\
\cline{2-6}
& (c) 
 & -10.119  &  -14.126  
 & -9.7256  &  -13.314
\\
\hline
{
12
}&  (a)
& -4.17 & -8.689
& -3.912 & -7.585
\\
\cline{2-6}
& (b)
& -4.056 &  -7.97
& -3.56 &  -6.5124
\\
\cline{2-6}
& (c) 
 & -46.88  &  -51.12  
 & -46.356  &  -49.52  
\\
\hline
\end{tabular}
\vskip 0.1in
\caption{Mass shifts of $\eta_c$ in magnetized nuclear
matter for densities of $\rho_0$ and 2$\rho_0$, asymmetric parameter,
$\eta$=0 and 0.5 and for magnetic fields, $eB/{m_\pi^2}$ as 4 and 12. 
These are tabulated without and with the contributions 
from anomalous magnetic moments of the nucleons 
in (a) and (b) respectively. The effects from the interaction
of spin to magnetic field (leading to mixing with $J/\psi$) 
are also taken into account in (c). }
\label{table2}
\end{table}

\begin{table}
\begin{tabular}{|c|c|c|c|c|c|}
\hline
\multicolumn{2}{|c|}{
} & \multicolumn{2}{|c|}{$ \eta=0 $} 
&  \multicolumn{2}{|c|}{$ \eta=0.5 $}\\
\cline{3-6}
\multicolumn{2}{|c|}
 {${eB}/{m_\pi^2}$}
 &  $  \rho_B=\rho_0 $ & $ \rho_B=2\rho_0 $ & 
$ \rho_B=\rho_0 $ & $ \rho_B=2\rho_0 $ \\
\hline
{
4}& (a) 
& -6.3735 &  -16.626
& -5.906 &  -14.286
 \\
\cline{2-6}
& (b) 
& -6.178 &  -15.265
& -5.453 & -13.35
 \\
\hline
{
12
}&  (a)
 & -6.49 &  -17.7
& -5.906 &  -14.286
\\
\cline{2-6}
& (b)
& -6.21 &  -15.169
& -5.218 &  -11.6264
\\
\hline
\end{tabular}
\vskip 0.1in
\caption{Mass shifts of $\chi_{c0}$ in magnetized nuclear
matter for densities of $\rho_0$ and 2$\rho_0$, asymmetric parameter,
$\eta$=0 and 0.5 and for magnetic fields, $eB/{m_\pi^2}$ as 4 and 12. 
These are tabulated (a) without and (b) with the anomalous magnetic
moments of the nucleons taken into consideration.}
\label{table3}
\end{table}

\begin{table}
\begin{tabular}{|c|c|c|c|c|c|}
\hline
\multicolumn{2}{|c|}{
} & \multicolumn{2}{|c|}{$ \eta=0 $} 
&  \multicolumn{2}{|c|}{$ \eta=0.5 $}\\
\cline{3-6}
\multicolumn{2}{|c|}
 {${eB}/{m_\pi^2}$}
 &  $  \rho_B=\rho_0 $ & $ \rho_B=2\rho_0 $ & 
$ \rho_B=\rho_0 $ & $ \rho_B=2\rho_0 $ \\
\hline
{
4}& (a) 
& -9.592 &  -24.34
& -8.894 &  -21.37
 \\
\cline{2-6}
& (b) 
 & -9.3  &  -22.824
 & -8.217  &  -20
  \\
\hline
{
12
}&  (a)
& -9.767 &  -25.451
& -8.894 &  -21.37
\\
\cline{2-6}
& (b)
& -9.343 &  -22.68
& -7.866 &  -17.448
\\
\hline
\end{tabular}
\vskip 0.1in
\caption{Mass shifts of $\chi_{c1}$ in magnetized nuclear
matter for densities of $\rho_0$ and 2$\rho_0$, asymmetric parameter,
$\eta$=0 and 0.5 and for magnetic fields, $eB/{m_\pi^2}$ as 4 and 12. 
These are tabulated (a) without and (b) with the anomalous magnetic
moments of the nucleons taken into consideration.}
\label{table4}
\end{table}

In the chiral effective model, 
the calculations are done in the mean field approximation.
In this approximation, the meson fields are treated as classical fields.
In the presence of a magnetic field, the proton has 
contributions from the Landau energy levels.
The effects of the anomalous magnetic moments of the
nucleons are also taken into consideration in the
present work of investigation of the in-medium masses
of the S- and P-wave charmonium ground states using the QCD sum rule approach.
For given values of baryon
density, $\rho_B$, isospin asymmetry, $\eta=(\rho_n-\rho_p)/(2\rho_B)$
($\rho_p$ and $\rho_n$ are the number densities for the proton
and neutron respectively), and magnetic field, 
the coupled equations of motion for the scalar fields,
$\sigma$, $\zeta$, $\delta$ and $\chi$ 
are solved. The  twist-2 and the scalar gluon condensates,
are calculated from the equations (\ref{g2approx})
and (\ref{chiglu}), which are then used to calculate
the values of $\phi_b$ and $\phi_c$
(given by equations (\ref{phib}) and (\ref{phicglu})
respectively). Using the values of $\phi_b$ and $\phi_c$, the
in-medium charmonium masses are calculated
using equation (\ref{masscharm}). For the S-wave charmonium
states, the effects from spin-magnetic field interaction are also
considered to calculate the effective masses for 
$J/\psi$ and $\eta_c$, using equation (\ref{effmass_swave}).

In isospin symmetric nuclear medium, at baryon densities, $\rho_{B} = 0$ 
and $\rho_{0}$, the values of the dilaton field, $\chi$ are
$409.76$ and $406.38$ MeV respectively and hence using equation 
(\ref{chiglu}), the values of the scalar gluon condensate 
$\left\langle \frac{\alpha_{s}}{\pi} G_{\mu\nu}^{a} {G^a}^{\mu\nu} 
\right\rangle$ turn out to be $(373$MeV$)^{4}$ and 
$(371.6$MeV$)^{4}$ for densities $\rho_B=0$ and $\rho_0$ 
respectively. The values of $\phi_{b}$,
turn out to be $2.3\times 10^{-3}$ and $2.27 \times 10^{-3}$ 
in the vacuum and at nuclear matter saturation 
density, $\rho_{0}$, respectively. These may be compared with
the values of $\phi_{b}$ to be equal to $1.8 \times 10^{-3}$ and 
$1.7 \times 10^{-3}$ respectively for $\rho_B=0$ and for 
$\rho_B=\rho_0$, obtained from the values of scalar gluon condensate 
of $(350 \;{\rm {MeV}})^{4}$ and $(344.81 \;{\rm {MeV}})^{4}$ respectively 
in vacuum and at nuclear saturation density, $\rho_{0}$
in Ref. \cite{klingl} in the linear density approximation. 
We might note here that the value of nuclear matter saturation 
density used in the present calculations is $0.15$ fm$^{-3}$ and 
in Ref. \cite{klingl}, it was taken to be $0.17$ fm$^{-3}$. 

In the present investigation, we choose $\xi=1$ for
study of the in-medium masses of $J/\psi$ and $\eta_c$, 
leading to the $\xi$ dependent coupling constant and charm quark mass,
$\alpha_s=0.21$ and $m_c=1.24$ GeV respectively
\cite{klingl,rein,kimlee}.
With these parameters and with $\phi_b$ calculated within the 
chiral effective model, the vacuum masses (in MeV) of $J/\psi$ and
$\eta_c$ are obtained to be 3196.2  and  3066.56 
respectively. 
In figures \ref{mjpsi_mag_qsr} and \ref{metac_mag_qsr},
the masses of the $J/\psi$ and $\eta_c$ are plotted
for isospin symmetric as well as asymmetric nuclear matter 
(with $\eta$=0.5) for typical values of the magnetic field
accounting for the anomalous magnetic moments (AMM) 
of the nucleons.
These results are compared to the case when the AMM effects
are not taken into consideration (shown as dotted lines).
The values of the mass shifts of $J/\psi$ and $\eta_c$ 
are presented in Tables 1 and 2 respectively,
for the cases of (a) without AMM, (b) with AMM
and (c) with AMM along with the effect of the 
spin--magnetic field interaction.
In isospin symmetric nuclear matter, at the nuclear matter 
saturation density, the values 
of mass shifts for $J/\psi$ and $\eta_{c}$ mesons, 
for $eB$ of $4m_\pi^2$ ($12m_\pi^2$)
are observed to be -4.3 (-4.316) and -4.07 (-4.056) MeV
when the AMM effects of nucleons are taken into account.  
These values are very similar to the zero magnetic field
values of -4.43 and -3.8 MeV respectively within 
the QCD sum rule calculation with gluon condensates
calculated from the chiral effective model \cite{amarvjpsi_qsr}, 
as well as the mass shifts of 
$-7$ MeV and $-5$ MeV respectively obtained in the linear
density approximation in Ref.\cite{klingl}. 
In Ref.\cite{kimlee} the operator product expansion was carried out upto 
dimension six and mass shift for $J/\psi$ was found to be $-4$ MeV at 
nuclear matter saturation density $\rho_{0}$. 
The difference in the masses of $J/\psi$ and $\eta_c$,
when the AMM effects are taken into account and
when these are not considered, is observed to be smaller
for the isospin asymmetric nuclear matter as compared
to the symmetric nuclear matter case. 

To study the medium modifications of the P-wave charmonia ($\chi_{c0}$
and $\chi_{c1}$), the value of $\xi$ is chosen to be 2.5,
for which the values of the coupling constant, $\alpha_s$=0.1948 
and $m_c$= 1.219 GeV \cite{rein,Song_Lee_Morita_Pwave_charmonia}.
The vacuum values for the scalar meson, $\chi_{c0}$ and
the axialvector meson $\chi_{c1}$ turn out to be
3720.6264 and 3877.77 MeV respectively.
The density dependence of the masses for these P-wave states, 
$\chi_{c0}$ and $\chi_{c1}$ are illustrated in figures 
\ref{mscalar_mag_qsr_xi2.5} and 
\ref{maxialvec_mag_qsr_xi2.5} for magnetic fields,
$eB$ as $4 m_\pi^2$ and $12 m_\pi^2$, for the isospin
symmetric nuclear matter ($\eta$=0) as well as for the extreme
isospin asymmetric matter ($\eta$=0.5).
The mass shifts of these mesons for the cases 
of accounting for the AMM effects of the nucleons are
given in tables III and IV respectively, for given densities,
magnetic fields and isospin asymmetry.
The mass drop obtained from QCD sum rule approach
are observed to be larger for the P wave charmonia
as compared to the S-wave charmonia. As may be seen from the
tables \ref{table1} -- \ref{table4}, 
in isospin symmetric nuclear matter, 
at nuclear matter saturation density,
for $eB=4 m_\pi^2$, the mass shifts (in MeV) 
for $J/\psi$, $\eta_c$, $\chi_{c0}$ and $\chi_{c1}$,
as calculated from the medium modifications of the
scalar and twist--2 gluon condensates,
within the QCD sum rule calculation (from equation (\ref{masscharm}) 
are observed to be -4.3, -4.1, -6.2 and -9.3 respectively,
when the AMMs of the nucleons are taken into account. 
These values are modified to
-9.9, -8, -15.26 and -22.82 at $\rho_B=2 \rho_0$ and
to -16.926, -12.23, -31.3294 and -39.74 at $\rho_B=4 \rho_0$.

There are dominant modifications to the masses of the 
S-wave charmonium states, $J/\psi$ and $\eta_c$,
in the presence of the spin-magnetic field  interaction,
leading to mass shifts (in MeV) of +1.75 and -10.12 for 
$J/\psi$ and $\eta_c$
respectively for $eB=4 m_\pi^2$ and +38.5 and -46.9 for
$eB=12 m_\pi^2$ at $\rho_B=\rho_0$ in isospin symmetric nuclear
matter. The modification of the mass
of $J/\psi$ at high magnetic fields should manifest 
in the reduction of the dilepton pairs produced from
$J/\psi$, for example at LHC.
The mixing of the state $\psi'$ with $\eta_c'$ has also been
considered along with the $J/\psi$--$\eta_c$ mixing in the presence 
of a magnetic field and the effects of magnetic field on the
formation times of these charmonium states were studied 
in Ref. \cite{Suzuki_Lee_2017}.
The formation times for the vector charmonium states
($J/\psi$ and $\psi'$) were observed to be delayed
as compared to those for their pseudoscalar partners
$\eta_c$ and $\eta'_c$. 
The mixing of $J/\psi (\psi')$ to $\eta_c (\eta'_c)$ 
might appear as the anomalous production of dileptons from 
$\eta_c(\eta'_c)$, in addition to $J/\psi (\psi')$ decaying
to dilepton pairs 
\cite{charmonium_mag_QSR,charmonium_mag_lee,Suzuki_Lee_2017}.
These mass modifications of the S-wave charmonium states,
arising due to the interaction of the
spin to magnetic field  should manifest in observable 
like dilepton spectra
in the high energy heavy ion collision experiments.

\section{Summary}
\label{sec:5}
In summary, in the present work, we have studied the 
mass modifications of the S-wave charmonium states, 
$J/\psi$ and $\eta_{c}$, and P-wave states, $\chi_{c0}$ and
$\chi_{c1}$, 
in the nuclear medium in the presence of strong magnetic fields,
using QCD sum rule approach. The mass modifications arise 
due to modifications of the scalar gluon condensate and twist-2
tensorial gluon operator, calculated within a chiral effective model.
The scalar and twist-2 gluon condensates in the nuclear medium
are obtained from the medium modification 
of a scalar dilaton field, $\chi$, which is introduced
in the chiral effective model to simulate the broken scale
invariance of QCD. The effects of anomalous magnetic moments 
of the nucleons are observed to be appreciable at higher densities
and higher magnetic fields. The mass shifts of the S-wave 
charmonium states are also calculated due to the interaction
of the spin to the magnetic field, and this modification
is observed to dominate over the mass shifts calculated within
the QCD sum rule aprroach, due to the modification
of the gluon condensates. The mixing of the vector $J/\psi$
with pseudoscalar $\eta_c$ should manifest in the dilepton spectra
of the high energy heavy ion collision experiments.

\acknowledgements
One of the authors (AM) is grateful to ITP, University of Frankfurt,
for warm hospitality and acknowledges financial support from Alexander
von Humboldt foundation when this work was initiated. 
 

\end{document}